# Van der Waals Epitaxy on Freestanding Monolayer Graphene Membrane by MBE

Jason Lapano[1], Ondrej Dyck[2], Andrew Lupini[2], Wonhee Ko[2], Haoxiang Li[1], Hu Miao[1], Ho Nyung Lee[1], An-Ping Li[2], Matthew Brahlek[1], Stephen Jesse[2], Robert G. Moore[1,*]

[1]*Materials Science and Technology Division, Oak Ridge National Laboratory, Oak Ridge, TN 37831, USA*

[2]*Center for Nanophase Materials Science, Oak Ridge National Laboratory, Oak Ridge, TN 37831, USA*

*Correspondence should be addressed to moorerg@ornl.gov

Abstract: Research on two-dimensional materials has expanded over the past two decades to become a central theme in condensed matter research today.  Significant advances have been made in the synthesis and subsequent reassembly of these materials using mechanical methods into a vast array of hybrid structures with novel properties and ever-increasing potential applications.  The key hurdles in realizing this potential are the challenges in controlling the atomic structure of these layered hybrid materials and the difficulties in harnessing their unique functionality with existing semiconductor nanofabrication techniques.  Here we report on high-quality van der Waals epitaxial growth and characterization of a layered topological insulator on freestanding monolayer graphene transferred to different mechanical supports.  This "templated" synthesis approach enables direct interrogation of interfacial atomic structure of these as-grown hybrid structures and opens a route towards creating device structures with more traditional semiconductor nanofabrication techniques.

The discovery in 2004 of a reliable and reproducible method to create graphene from mechanical exfoliation created an avalanche of research activity in both academic and industrial arenas.[1] The field quickly expanded beyond graphene into a vast array of other two-dimensional materials with potential applications ranging from electronics and optoelectronics to catalysis, sensing, energy generation and storage.[2-5] Many of these two-dimensional atomic crystals form layered structures with strong in-plane covalent bonds and neighboring layers held together with relatively weak van der Waals forces creating a dreamscape of atomic building blocks stacked together for novel functionalities.[6] Advances in mechanical exfoliation and the use of graphene for remote epitaxy and delamination of material layers has demonstrated application potential but significant challenges still remain.[7,8] The ability to fabricate, characterize and ultimately control material structures and devices that are only a few atoms thick is a primary hurdle between novel properties and scalable applications using existing semiconductor nanofabrication techniques such as lithography and etching.

Graphene has also been used as suitable substrate for the van der Waals epitaxial growth of a wide range of quantum materials by Molecular Beam Epitaxy (MBE). New insights into the unconventional superconductivity of FeSe, dimensional effects in charge density waves, the transition from indirect to direct bandgap in transition metal dichalcogenides and the quantum spin Hall state in 1T'-WTe$_2$ are made possible by the underlying support of a graphene substrate.[9-13] The unique properties of graphene make it a robust template for the growth of ultra-thin and two-dimensional nanomaterials using MBE but existing graphene substrates are typically formed from repeated high-temperature flashing of SiC.[14] Hence, despite the wide use of MBE in industry, similar challenges remain for the MBE grown two-dimensional materials in the characterization and device applications using existing technologies. Here we report on the van der Waals epitaxial growth of a layered topological insulating material directly on monolayer graphene transferred to an amorphous support and a freestanding monolayer graphene membrane extended over micrometer scale holes with subsequent characterization of the as-grown material properties. This "templated" synthesis approach can help bridge the gap between van der Waals materials and traditional semiconductor nanofabrication techniques as well as provide a novel way to directly interrogate the atomic and interfacial structure of low-dimensional materials grown by MBE with a Scanning Transmission Electron Microscope (STEM).

Graphene is a two-dimensional honeycomb-like arrangement of carbon atoms with a thickness of a single atom.[15] The carbon atoms form strong sp$^2$ hybridized bonds with subsequent layers linked by weaker van der Waals forces. The atomic configuration generates an unusual electronic band structure with massless Dirac electrons and ballistic transport making it possible to investigate quantum effects at room temperature.[16,17] Beyond the fundamental interest in its quantum properties, graphene has been shown to be 200 times stronger than steel and can endure elastic deformation on the order of 20% without breaking.[18,19] Its robust structural integrity enables its isolation through destructive methods such as exfoliation, arc discharge and unzipping of carbon nanotubes.[1,20,21] However, methods for the synthesis of graphene through the assembly of carbon atoms have been developed in the form of epitaxial growth, gas-phase and organic synthesis in addition to Chemical Vapor Deposition (CVD).[22-25] The CVD synthesis method, combined with its robust strength and flexibility, has enabled the creation of graphene on different mechanical supports with the ability to be transferred to other surfaces.[26] The process has become standardized to the point that graphene transferred to thermally grown SiO$_2$ on Si wafers and holey Si$_3$N$_4$ Transmission Electron Microscope (TEM) grids are now commercially available. This raises the question as to whether these transferred graphene composite systems can be used for a

"templated growth" approach for the epitaxial synthesis of quantum materials. Towards this goal we demonstrate epitaxial growth of the topological insulator $Bi_2Se_3$ by MBE directly onto monolayer graphene transferred to $SiO_2$/Si substrates ($Bi_2Se_3$/G/SiO) and holey $Si_3N_4$ TEM grids ($Bi_2Se_3$/G/hSiN).

Thin films of $Bi_2Se_3$ were chosen for study as it is a prototypical topological insulator with a bulk bandgap $\Delta \sim 300$ meV and topologically protected in-gap interfacial electronic states consisting of a Dirac cone appearing at the material-vacuum boundary.[27] These surface states have been characterized in detail previously using techniques as angle resolved photoemission spectroscopy (ARPES) and scanning tunneling microscopy/spectroscopy (STM/STS).[28-35] The two dimensional, spin-momentum locked electronic states with topological protection makes them suitable for spin devices, low power electronics and fault tolerant quantum computing applications.[36-38] In addition, more exotic states such as the quantum anomalous hall effect and topological superconductivity demonstrating non-Abelian statistics can be created through proximity induced effects when heterostructures are created with magnetic or superconducting materials.[39-41] Finding new pathways to create and interrogate hybrid topological structures is critical towards harnessing exotic quantum states for potential applications.

Commercial graphene transferred to different mechanical supports were purchased from different vendors (Ted Pella, ACS Material) and cleaned to remove residual photoresist prior to growth using a multi-step process. The graphene is first annealed in a flow of forming gas (5% Hydrogen, 95% Nitrogen) at 350 °C for approximately 12 hours.[42] The graphene plus amorphous mechanical support are then glued to sample plates using silver epoxy and annealed in vacuum at 400-450 °C for 48 hours prior to growth. After cleaning, samples were transferred to a home-built molecular beam epitaxy system operating at a base pressure of $<5\times10^{-10}$ Torr for $Bi_2Se_3$ deposition. The samples were annealed to 600 °C before growth for a final surface cleaning and to ensure uniformity in growth conditions. Elemental Bi and Se was supplied via thermal effusion cells and fluxes were calibrated using a quartz crystal microbalance. Bi was supplied at a flux of $2\times10^{13}$ $cm^{-2}s^{-1}$ and to suppress formation of Se defects Se was supplied in excess at a flux of $10\text{-}20\times10^{13}$ $cm^{-2}s^{-1}$. The samples were cooled to 135 °C where a 3 quintuple layer (QL) buffer layer was deposited to aid in nucleation.[43] Samples were then heated to 235 °C where the remaining 17 QLs were deposited for a final film thickness of 20 nm.

To interrogate the quality of the $Bi_2Se_3$ films grown using this templated approach and to characterize the topological surface states, films were transferred *in situ* to a separate ARPES chamber for mapping the electronic structure. The ARPES system consists of a Scienta DA30L hemispherical electron analyzer maintained at a base pressure of $< 5\times10^{-11}$ Torr with a base sample temperature T < 10 K and a He lamp (He-I$\alpha$ = 21.2 eV) light source. Figure 1 summarizes the ARPES results on the $Bi_2Se_3$/G/$SiO_2$ and $Bi_2Se_3$/G/hSN films compared to the film grown on a more traditional single crystal $Al_2O_3$ substrate. The band structure of the film grown on the $Al_2O_3$ substrate is shown in Figure 1a and is similar to previous results.[28, 29] The linearly dispersing Dirac cone is the signature of the topological insulating state and is thus the focus of our investigation. Figures 1b and 1c show the resulting topological band structure of $Bi_2Se_3$/G/SiO and $Bi_2Se_3$/G/hSiN, respectively. The topological band structure observed on the $Bi_2Se_3$/G/SiO film is nearly identical to that observed on the $Al_2O_3$ substrate observed in Figure 1a and to films grown on bi-layer graphene formed on the surface of SiC.[35] However, the Dirac cone for the $Bi_2Se_3$/G/hSiN shown in Figure 1c is broader than that observed from the other films. As will be discussed later, this broad and "fuzzy" electronic structure is likely due to a larger surface roughness of the TEM grid rather than inferior crystalline quality of the film itself. The insets in Figure 1a-c show constant energy maps of the bulk electronic bands near the center of the Brillouin zone at a Binding Energy $E_B$ = -0.58 eV below the Fermi level.  The film grown on $Al_2O_3$ clearly shows the 6-fold symmetry of the bulk bands in the first Brillouin zone as expected for twinned crystals of 3-fold symmetry.  If the domains of the films grown on graphene were randomly oriented, one would expect a ring of intensity from the bulk bands due to the lack of long-range order.  As can be seen in the Figure 1b and 1c insets, lobes of intensity around $k_x$ = ± 0.5 (1/Å) can still be observed, indicating long range order over the ARPES beam spot size (~1 mm).  The multiple copies of lobes in Figure 1b and broader lobes in Figure 1c suggest some disorder in the orientation of the domains but they are not random and have a limited number of specific domain orientations indicative of van der Waals epitaxial film growth with some texturing.

For investigating the atomic and electronic structure, the $Bi_2Se_3$ film samples were transferred from the MBE chamber to an STM chamber via an UHV vacuum suitcase (P < $1x10^{-10}$ Torr) to avoid atmospheric exposure. Room temperature STM/STS data were taken with Omicron VT-STM that operated at a base pressure P < $1x10^{-10}$ Torr. Figures 2a and 2b shows the STM topographic images of $Bi_2Se_3$/G/SiO and $Bi_2Se_3$/G/hSiN, respectively. The $Bi_2Se_3$ films grown on transferred monolayer graphene have a morphology similar to those grown on crystalline substrates as has been reported previously.[31, 43]  Still, as compared to the large flat islands seen on $Bi_2Se_3$/G/SiO, the $Bi_2Se_3$/G/hSiN displays smaller terraces, larger corrugation, and increased number of defects, likely due to the increased corrugation and holes in graphene on holey $Si_3N_4$ as also shown in the STEM images in Figure 3. On the clean regions of $Bi_2Se_3$ film, we can observe atomic lattices of the topmost Se layer (Figure 2c), confirming the film quality. In addition, the STM image shows interesting native defects such as screw dislocations and grain boundaries in the absence of external contamination, suggesting these sample are a good platform to investigate novelties in the electronic structure associated with such native defects.[44, 45]  The dI/dV spectra, a measure of the local density-of-states (LDOS),  were taken by conventional lock-in method with modulation frequency f = 1 kHz and modulation amplitude V = 10 mV. The dI/dV spectra taken on the flat part of the sample shows typical V-shape of topological insulators (Figure 2d).[34, 46] Since the LDOS minimum corresponds to the Dirac point, we can determine the Dirac point of $Bi_2Se_3$ film to be V = -0.3 eV, which is consistent with the ARPES results and previous reports.[47]

To better understand the orientation of the film domains X-ray diffraction was performed using a 4-circle goniometer and a monochromate Cu-$K_\alpha$ source with a beam footprint of 1 mm × 5 mm.  Off-axis [*H* 0 *L*] peaks are used to investigate the alignment of the $Bi_2Se_3$ domains relative to the substrates. Figure 2e shows a polar plot of the [1 0 10] reflection of $Bi_2Se_3$ as well as the $Al_2O_3$ [1 0 4] versus the azimuthal angle (φ). Here it can be seen that the $Al_2O_3$ [1 0 4] exhibits the expected 3-fold rotational symmetry. In contrast, $Bi_2Se_3$ shows 6 peaks separated by 60° indicative of epitaxial twinning due to two equivalent orientations for nucleation on the $Al_2O_3$ c-plane surface, as schematically shown in the inset of Figure 2e. As shown in Figure 2f, the azimuthal scan or the $Bi_2Se_3$/G/hSiN sample reveals that the films are textured. Like $Al_2O_3$, graphene exhibits a triangular surface structure and therefore, one would expect a similar nucleation of $Bi_2Se_3$ showing a 60° angular separation of the [1 0 10] reflections. Grouping sets of peaks separated by integer units of 60° reveals there are predominately 4 sets of peaks with relative orientation of 0°, 17°, 30°, and 51°. These orientations reflect the predominant grain orientations of the graphene. However, directly quantifying the grain texture is not possible using X-ray diffraction for a single graphene monolayer.  The diffraction data for the films grown on the monolayer graphene reveal some disorder in the orientation of the $Bi_2Se_3$ domains but these results reveal that the domains are not randomly oriented with a limited number of preferred orientations. The domain structure revealed by STM (Figure 2a) and STEM (Figure 3c) also show large triangular domains with a similar grain size to those observed on $Al_2O_3$ but with a limited number of domain orientations.[31, 43]

To examine the $Bi_2Se_3$/G/hSiN film structure High-Angle Annular Dark Field (HAADF)-STEM imaging was performed on the as-grown samples, as shown in Figures 3 and 4. Images were acquired using a Nion UltraSTEM 200 operated at 100 kV accelerating voltage with a nominal beam current of 10 pA and convergence angle of 30 mrad. It should be emphasized that the single atom thick layer of graphene extends over the 2 μm diameter holes in the $Si_3N_4$ TEM grid, creating a freestanding graphene membrane, on which the $Bi_2Se_3$ is grown. Figure 3a and 3d show representative overview images of the $Bi_2Se_3$/graphene stack comparing the role of graphene holes and tears. In the absence of the graphene support, the $Bi_2Se_3$ layer terminates. There is a variation in the quality of the graphene from grid to grid and Figure 3d shows growth across a continuous graphene layer without holes or tears. In this case a smooth $Bi_2Se_3$ film grows across the full aperture with fewer defects observed. Figure 3a-c shows a magnified view of a defect region where $Bi_2Se_3$ has not fully covered the graphene and we are able to visualize the growth front. A region with a single layer of $Bi_2Se_3$ is shown in the Figure 3c inset. The heavier Bi atoms appear brighter in HAADF-STEM imaging due to their heavier mass which results in the appearance of the brighter hexagons while the dimmer Se atoms can be seen at the center of the hexagons.[48] Figure 3e and 3f shows a higher magnification images of the $Bi_2Se_3$ on a region of continuous graphene with Figure 3f showing an atomically resolved image of the continuous film. The Fourier transform of the Figure 3f inset reveals the hexagonal symmetry of the crystalline van der Waals epitaxial thin film.

We examine the growth termination of the $Bi_2Se_3$ around graphene defects more closely in Figure 4. Figure 4a shows a region of the $Bi_2Se_3$ film where the underlying graphene was broken. We can see that the $Bi_2Se_3$ film terminates at the edge of the hole with intensity steps indicative of different layer thickness. In Figure 4b-d we examine another region with the $Bi_2Se_3$ film growth up to the graphene edge. Figure 4c shows a magnified view of this edge termination where the underlying graphene can be seen protruding beyond the $Bi_2Se_3$ film by a few angstroms. Figure 4d shows an intermediate magnification view of this edge taken from the boxed area in 4b with a Fourier transform of the boxed area illustrating the high-quality hexagonal crystal. Our observations reveal the van der Waals epitaxial growth of large domains of $Bi_2Se_3$ is possible up to the edge of the graphene defects.

The combined results demonstrate high quality van der Waals epitaxial $Bi_2Se_3$ grown on the transferred monolayer graphene. The STM plus STEM results indicate there is less uniformity in the quality of the film over large regions of the graphene on the TEM grids. The larger corrugations observed due to the morphology of the underlying holey $Si_3N_4$ as well as rips and tears in the graphene extended over the TEM grid holes generate defects in the films. These defects appear as bright spots in both the STM and STEM images with the number varying across the sample and from sample to sample. The increased roughness of the $Si_3N_4$ and growth defects result in a less defined electron ejection angle during the photoemission process, which yields the lower quality "fuzzy" ARPES spectra noted earlier. However, the STEM images indicate the defect density is limited to some portions of the sample and the fact that clear ARPES is observed, even with a ~1 mm spot size, is quite remarkable. The observed lobes in the bulk bands, as seen from ARPES, combined with the limited number of domain orientations observed by XRD indicates highly oriented domains over macroscopic length scales despite the expected wrinkles, folds and dislocations in the underlying graphene that can change the lattice orientation and trigger growth defects.[5] As improvements are made to the quality of the graphene transferred to different mechanical supports we anticipate further reduction in the film texturing and longer range van der Waals registry between the monolayer graphene and film overlayer with improved domain alignment.

The freestanding monolayer graphene membrane extended over 2 µm holes in the TEM grids provides a unique perspective to investigate the film growth. The interrogation of an as-grown epitaxial film with the STEM electron beam normal to the film surface provides new opportunities to investigate interfacial structure and native defects at heterojunction interfaces of dissimilar materials. Such information can be used to optimize growth conditions of heterojunctions and better understand the connections between native defects and observed properties. As seen in Figure 3e, the film grows up to the edges of the holes and tears in the graphene membrane and maintains its crystalline structure at the vacuum boundary. This suggests it may be possible to pattern graphene on $SiO_2$ using traditional nanofabrication techniques for the growth of films in the specific geometric patterns necessary for device fabrication. In addition, ultrathin two-dimensional materials may also be directly shaped and manipulated using the STEM beam as has been demonstrated for graphene.[49-51] These observations show the possibility of bridging the gap between two-dimensional MBE material synthesis and existing nanofabrication techniques through a templated synthesis with graphene transferred to patterned structures as the substrate.

In summary we have demonstrated van der Waals epitaxial growth of the prototypical topological insulator $Bi_2Se_3$ on monolayer graphene transferred to different amorphous mechanical supports. Characterization of the film's atomic and electronic structure using several techniques reveals that the resulting films are high quality over macroscopic length scales. The amorphous $SiO_2$ and holey $Si_3N_4$ TEM grids used in this study demonstrate the potential of using transferred graphene as an epitaxial template for the growth of a wide array of quantum materials. The epitaxial growth of a film over a 2 µm monolayer graphene membrane opens new opportunities for the investigation and control of materials demonstrating exotic quantum phenomena.

The authors would like to acknowledge fruitful discussion with Felix Luepke. The MBE film synthesis, ARPES and STM work were supported by the U. S. Department of Energy (DOE), Office of Science, National Quantum Information Science Research Centers. The STM characterization was conducted at the Center for Nanophase Materials Sciences, which is a DOE Office of Science User Facility. Electron microscopy and portions of the ARPES efforts were supported by the U.S. DOE, Office of Science, Basic Energy Sciences, Materials Sciences and Engineering Division.

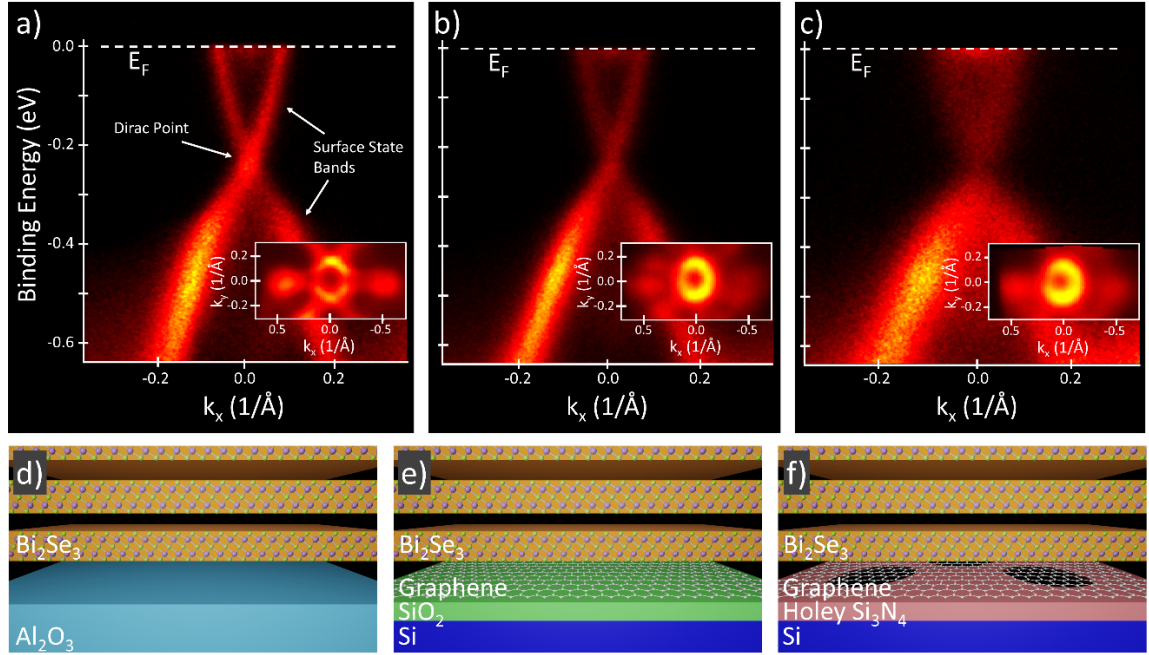

Figure 1. Topological electronic band structure of $Bi_2Se_3$ films. (a-c) ARPES characterization of topological surface bands at Brillouin zone center of $Bi_2Se_3$ films grown on $Al_2O_3$, $Bi_2Se_3$/G/SiO and $Bi_2Se_3$/G/hSiN respectively. Insets are constant energy map at $E_B$ = -0.58 eV for the respective samples. (d-f) Schematic diagrams of the multi-layered $Bi_2Se_3$ films grown on the different substrates corresponding to ARPES data shown in (a-c).

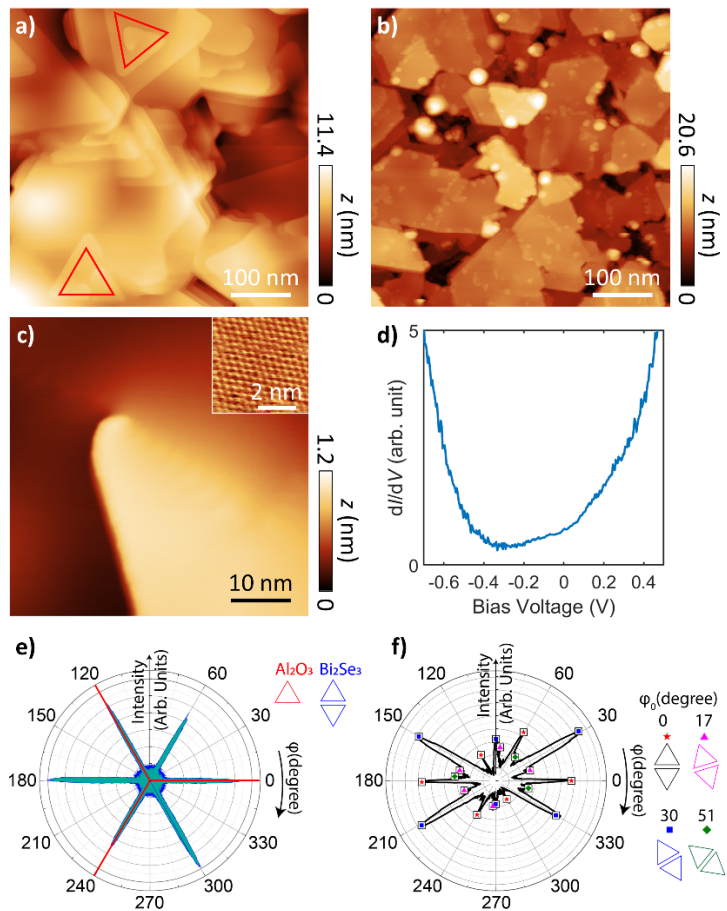

Figure 2. Bi$_2$Se$_3$ film morphology and structure. (a) Large-scale room temperature STM image of the Bi$_2$Se$_3$/G/SiO sample (V = 2 V, I = 1 pA). Red triangles highlight different domain orientations. (b) Large scale STM image of the Bi$_2$Se$_3$/G/hSiN sample (V = 2 V, I = 1 pA). (c) An STM image of the area with screw dislocation and grain boundary in Bi$_2$Se$_3$/G/SiO sample (V = 1 V, I = 1 pA). Inset shows a zoomed in image of the flat area with atomic resolution (V = 0.3 V, I = 300 pA). (d) dI/dV spectrum taken on the flat area of (c). The minimum LDOS at V = -0.3 V corresponds to the Dirac point. (e-f) X-ray diffraction azimuthal φ scans, diffraction intensity versus φ angle shown in radial plots for the [1 0 10] Bi$_2$Se$_3$ and the [1 0 4] Al$_2$O$_3$ family of peaks (e), and about the [1 0 10] Bi$_2$Se$_3$ peaks for the Bi$_2$Se$_3$/G/SiO sample (f).

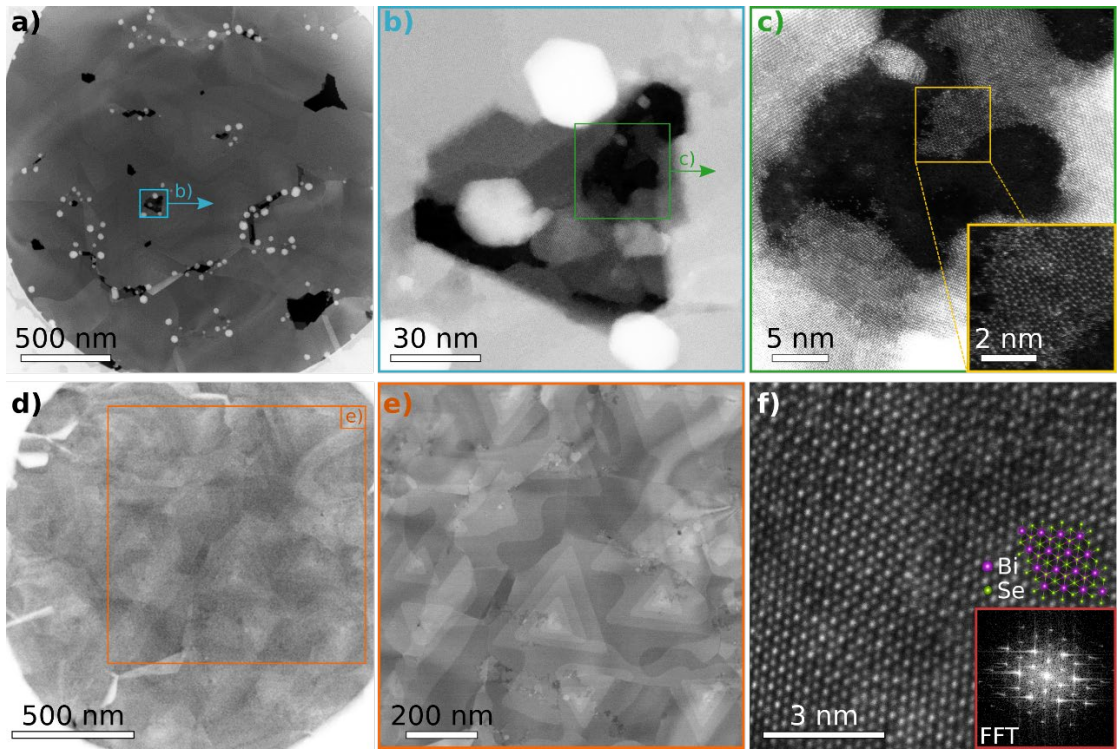

Figure 3. Summary HAADF-STEM images of $Bi_2Se_3$/G/hSiN. (a) Example overview image of $Bi_2Se_3$ on suspended graphene. In this region, the graphene had a few holes and defects. (b) Magnified view of the region marked in (a). (c) Magnified view of the region marked in (b). Here, the underlying graphene is unbroken and we can observe the edges of the growth front. Inset shows a magnified view of the single layer region. (d) Overview of the $Bi_2Se_3$ on an unbroken region of the graphene. (e) Magnified view of the region marked in (d). (f) Atomically resolved image of a $Bi_2Se_3$/graphene region. The Fourier transform of the image is shown inset, illustrating a strong hexagonal pattern. A ball and stick crystal structure for a single layer of $Bi_2Se_3$ is overlaid and aligned to the image.

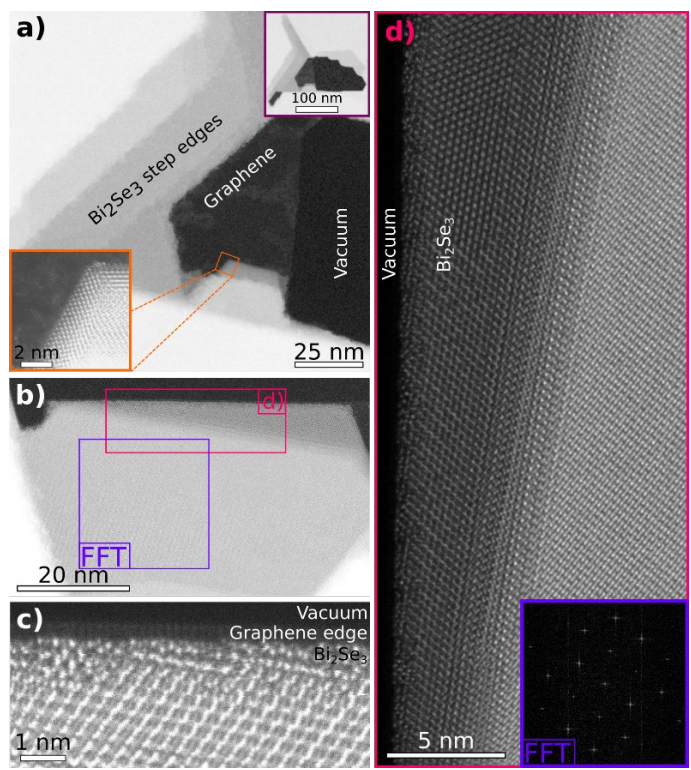

Figure 4. Examination of the Bi$_2$Se$_3$ edge termination. (a) HAADF-STEM image of a hole in the Bi$_2$Se$_3$ where a portion of the graphene is also broken. The Bi$_2$Se$_3$ grows right up to the graphene edge over a substantial portion of the hole (see upper right inset). Growth layers are seen in the Bi$_2$Se$_3$ step edges. A magnified view of a growth facet on the graphene surface is shown in the lower left inset. (b) Closer examination of a region where the Bi$_2$Se$_3$ has grown up to the graphene edge. (c) A magnified view of the edge where the underlying graphene can be seen protruding by several angstroms. (d) Larger view of the edge structure acquired in the location indicated in (b). Inset is a Fourier transform from the location indicated in (b) illustrating the hexagonal crystal periodicity.